**Dopant-segregation to grain boundaries controls electrical conductivity of n-type NbCo(Pt)Sn half-Heusler alloy mediating thermoelectric performance**

Ting Luo [a,*], Federico Serrano-Sánchez [b], Hanna Bishara [a], Siyuan Zhang [a], Ruben Bueno Villoro [a], Jimmy Jiahong Kuo [c], Claudia Felser [b], Christina Scheu [a], G. Jeffrey Snyder [c], James P. Best [a], Gerhard Dehm [a], Yuan Yu [d], Dierk Raabe [a], Chenguang Fu [b], Baptiste Gault [a, e, *]

[a] Max-Planck-Institut für Eisenforschung GmbH, Max-Planck Straße 1, 40237, Düsseldorf, Germany
[b] Max Planck Institute for Chemical Physics of Solids, Nöthnitzer Str. 40, 01187 Dresden, Germany
[c] Materials Science & Engineering (MSE), Northwestern University, Evanston, IL 60208, USA
[d] Institute of Physics IA, RWTH Aachen University, 52056 Aachen, Germany
[e] Department of Materials, Royal School of Mines, Imperial College, Prince Consort Road, London, SW7 2BP, UK

Keywords: NbCoSn, half-Heusler, grain boundary segregation, grain boundary resistivity, thermoelectric

# Abstract:

Science-driven design of future thermoelectric materials requires a deep understanding of the fundamental relationships between microstructure and transport properties. Grain boundaries in polycrystalline materials influence the thermoelectric performance through the scattering of phonons or the trapping of electrons due to space-charge effects. Yet, the current lack of careful investigations on grain boundary-associated features hinders further optimization of properties. Here, we study n-type NbCo$_{1-x}$Pt$_x$Sn half-Heusler alloys, which were synthesized by ball milling and spark plasma sintering (SPS). Post-SPS annealing was performed on one sample, leading to improved low-temperature electrical conductivity. The microstructure of both samples was examined by electron microscopy and atom probe tomography. The grain size increases from ~230 nm to ~2.38 μm upon annealing. Pt is found within grains and at grain boundaries, where it locally reduces the resistivity, as assessed by *in situ* four-point-probe electrical conductivity measurement. Our work showcases the correlation between microstructure and electrical conductivity, providing opportunities for future microstructural optimization by tuning the chemical composition at grain boundaries.

# 1. Introduction

Thermoelectric (TE) devices exploit the Seebeck effect to directly convert heat gradients into electricity [1]. This environmentally-friendly energy conversion approach harvests waste heat and is particularly attractive as an alternative or complement to non-renewable, fossil fuels for power generation [2]. In addition, some advanced TE modules have been developed to transform the human body heat to power wearable devices like electrocardiographic systems [3]. The conversion efficiency of TE materials is determined by the dimensionless figure-of-merit ($zT$). This parameter is expressed as $zT = \alpha^2 \sigma T/\kappa$, where $\alpha$, $\sigma$, $T$ and $\kappa$ are Seebeck coefficient, electrical conductivity, absolute temperature and total thermal conductivity, respectively [2, 4, 5]. The thermal conductivity $\kappa$ compounds the lattice and electronic thermal conductivities, $\kappa_L$ and $\kappa_e$ respectively [2, 4, 5].

Many promising materials have been developed for TE applications, such as $Bi_2Te_3$ [6, 7], $Mg_3Sb_2$ [8, 9], half-Hesulser (HH) [10], GeTe [11, 12] and so on. Among these materials, HH compounds, with a stoichiometry XYZ, stand out as state-of-the-art TE materials for medium-to-high temperature applications [10, 13, 14]. As the three constituents X, Y, Z can be individually substituted, there are abundant opportunities to optimize the TE performance of HHs by doping and isoelectronic alloying, i.e. tuning the alloy's composition [13, 15, 16]. HHs have an intrinsic high power factor in the order of ~$10^{-3}$ W m$^{-1}$ K$^{-2}$ [10, 15], due to the large optimal carrier concentration stemming from its heavy and flat band structure. The optimal power factor of HHs is 2-3 times higher than light-band PbTe [17].

However, the key challenge for the application of HHs is their intrinsically large thermal conductivity in the order of 10Wm$^{-1}$K$^{-1}$ at T=300K, i.e. an order higher than GeTe [11]. Many efforts have been made to reduce $\kappa_L$ by introducing structural defects, e.g. dislocations [18], nano-precipitates [19], and grain/phase boundaries [18, 20, 21]. For instance, Rogl et al. [18] conducted severe plastic deformation of high-pressure torsion on hot-pressed NbFeSb-based alloys. The thermal conductivity was reduced by ~60% from 75 to 28.5 mW cm$^{-1}$ K$^{-1}$ due to the grain refinement and a high concentration of deformation-induced defects of vacancies and dislocations. However, some defects such as grain boundaries (GBs) or phase boundaries can modify the local transport properties in a way that is not necessarily beneficial to zT due to the tradeoff between phonon scattering and electron transport [18, 22-24]. Thus, a net enhancement of zT requires meticulous control of the microstructure and a deep understanding of the local structural and chemical states of defects, as well as their influence on the transport processes.

The mean free path of charge carriers is found to be a few nm in HHs [25]. While it depends on the grain size, the spacing between GBs is typically much larger. GBs are thus expected not to scatter charge carriers and have no significant impact on electrical conductivity. However, recent reports on $Mg_3Sb_2$-based thermoelectric materials [9, 26] have shown that GB could act as a well of charge carriers and thus decreases the electrical conductivity of the material. A two-phase model is typically used to account for the influence of GB on the transport properties of the bulk material [26]. However, if this model assumes a structural discontinuity at GBs, the local chemical composition is typically ignored. As thermodynamically explained by the Gibbs adsorption isotherm, a chemically decorated interface is the rule and not the exception in 'real' microstructures [27]. The segregants can have either detrimental or beneficial effects on the microstructure evolution and properties [27]. Understanding the interaction between dopants and GBs is hence critical to interpreting the TE performance. Insights into the effect of GB composition on local transport properties provide extra freedom to control the TE performance by segregation engineering.

Here, we provide insights into GBs in high-performing n-type $NbCo_{1-x}Pt_xSn$ HH alloys to reveal the mechanisms leading to the enhanced TE performance. Pt replaces Co in the lattice with one more valence electron and is an effective dopant. Pt-doping leads to significant improvements in the electrical power factor and reduces lattice thermal conductivity, as reported for $NbCo_{1-x}Pt_xSn$ ($x$=0.00-0.15) [28]. This previous report showed an increased zT by post-annealing with increased grain size, but the local structural details of GBs and their correlation to TE properties were not investigated. We hence focus on the local chemical composition of GBs for $NbCo_{1-x}Pt_xSn$ samples with $x$=0.06 (not annealed) and $x$=0.05 (annealed) that show striking performance improvement to reveal the influence of GB chemistry on TE properties. Particularly, the electrical conductivity of $x$=0.05 is higher than the $x$=0.06 sample. The distribution of Pt dopants was characterized by atom probe tomography (APT), which is a burgeoning technique suited to relate nano-scale composition and TE performance, with high chemical sensitivity and 3D spatial resolution [29-32]. APT show that the intraranular Pt-content is similar for both alloys, indicating that the improvement in the electrical conductivity does not stem from the doping level, rather, from the evolution of microstructure. We reveal strong levels of Pt-segregation to GBs. *In situ* four-point-probe local electrical conductivity measurement was then conducted to reveal the impact of Pt enrichment at GBs, at a micrometer scale, on the GB-electrical conductivity. The present results deepen our understanding of the role of GBs on transport properties and provide a guide to further optimize the TE performance by tuning the chemical composition at GBs.

## 2. Experimental procedures

*2.1 Sample synthesis*

NbCoSn, $NbCo_{0.95}Pt_{0.05}Sn$ and $NbCo_{0.94}Pt_{0.06}Sn$ ingots were synthesized by arc-melting in an argon atmosphere. Nb slug, Co slug, Sn shot and Pt shot, were used as starting materials. The ingots were flipped and remelted three times for compositional homogeneity. Afterwards, the ingots were vacuum-sealed in quartz ampoules with Ta foils and annealed at 1073 K for 7 days. The bulk ingots were grounded and ball-milled (BM) using WC grinding balls (Pulverisette 7, Fritsch). The powders were then placed in graphite dies of 10 mm and consolidated into disks by spark plasma sintering (SPS). Post-SPS annealing was performed at 1073 K for 7 days on one NbCo$_{0.95}$Pt$_{0.05}$Sn sample.

In the following, undoped NbCoSn, doped NbCo$_{0.94}$Pt$_{0.06}$Sn and NbCo$_{0.95}$Pt$_{0.05}$Sn with post-SPS annealing will be referred to as NbCoSn, NbCoSn-Pt and NbCoSn-Pt-AN, respectively. *In situ* four-point-probe local electrical conductivity measurement was conducted on the pre-ball milled sample (i.e. arc melted and annealed sample) with a nominal composition of NbCo$_{0.95}$Pt$_{0.05}$Sn.

*2.2 Microstructure analysis*

The microstructure of Pt-doped NbCoSn, with and without post-SPS, was characterized using a Zeiss Merlin scanning electron microscope (SEM), operated at 30 kV in backscattered electron (BSD) mode. Energy-dispersive X-ray spectroscopy (EDX) was performed on the pre-BM sample in Zeiss Merlin SEM at 15kV and 6nA. Electron backscatter diffraction (EBSD) was conducted in Zeiss Sigma SEM at 15kV, 9nA and with a step size of 1μm. Further microstructural analysis was carried out using atom probe tomography (APT). APT needle-shaped specimens were prepared using dual-beam SEM/focused-ion-beam (FIB) instruments (Helios Nanolab 600i and Thermo Fisher Scios2), respectively following the procedures described elsewhere [33]. The final ion-milling step to minimize the beam damage was performed at 2 kV for the APT specimen. LEAP 5000XR instrument was operated in laser pulse mode at 50 K, 125 kHz laser frequency, and 30-35 pJ laser energy. Data reconstruction and processing were performed using the Cameca IVAS 3.8.4 software tool.

*2.3 Thermoelectric performance*

The Seebeck coefficient and electrical conductivity were measured simultaneously using an ULVAC ZEM-3 system. The diffusivity was determined by the laser-flash method (LFA 457, Netzsch) and the thermal conductivity was calculated as the product of thermal diffusivity, specific heat and density.

*2.4 Local electrical measurement*

A four-probe technique for local electrical measurement was conducted in Zeiss a scanning electron microscope (Gemini 500). The sample was inserted into SEM on a stage where four-electrode needles are held each by independent manipulators. The four needles were positioned in a linear configuration with equal spacing of ~5 μm, within the grain interior and crossing a high angle grain boundary. A series of 100 electrical current pulses (direct current, 10 mA) is applied to the two outmost needles with 10 ms long pluses using a current generator (Keithly 6221). Voltage was measured and averaged between the two middle needles at the pulse half-time by employing a nano voltmeter (Keithly 2182A). In the calculation of resistivity, the deviation in spacing between adjacent needles was considered and details are referred to elsewhere [34].

# 3. Results and discussion

*3.1 Thermoelectric performance*

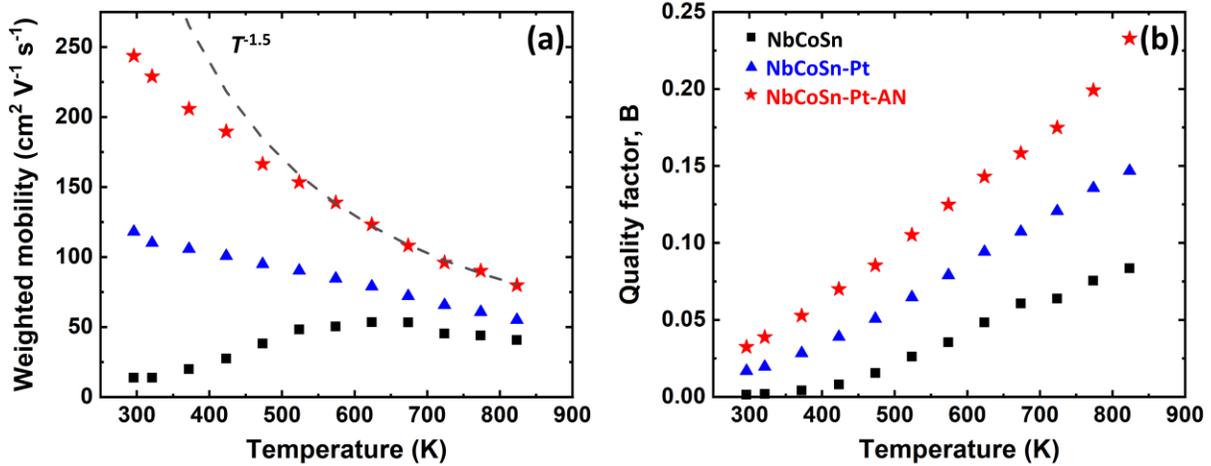

**Fig. 1.** Temperature dependence of (a) weighted mobility, and (b) quality factor B for NbCoSn, NbCoSn-Pt and NbCoSn-Pt-AN. The dash line in (a) represents the temperature dependence of acoustic-phonon scattering ($\mu_w \sim T^{-1.5}$).

In order to exclude the effect of different doping levels on transport properties, we calculated the weighted mobility ($\mu_w$) from the experimental electrical conductivity and Seebeck coefficient as $\mu_w$ is independent of the carrier concentration [35]. Fig. 1(a) shows the temperature dependence of $\mu_w$ for NbCoSn, NbCoSn-Pt and NbCoSn-Pt-AN. The raw measured transport properties were reported previously and are appended in Fig. S1 (Supplementary materials) [28]. The dash line in Fig. 1(a) represents the $\mu_w \sim T^{-1.5}$ relationship for the acoustic-phonon scattering of charge carriers [35]. The $\mu_w$ of pristine NbCoSn increases with temperature until 625K. This thermally activated behavior of $\mu_w$ indicates GB-scattering of charge carriers, which has also been found in other HH alloys [36] and $Mg_3Sb_2$ [26].

By adding Pt dopants, the $\mu_w$ of NbCoSn-Pt is increased, particularly at low temperatures. The increased $\mu_w$ can either result from the enhanced Hall carrier mobility ($\mu_H$) or the enlarged density-of-states effective mass ($m^*$), this latter hypothesis was previously dismissed [28]. Therefore, the increased $\mu_w$ only results from the increased $\mu_H$ as the GB-scattering of charge carriers is reduced by doping Pt. No thermally-activated $\mu_w$ is observed in NbCoSn-Pt, but its temperature dependence is much weaker than the pure acoustic phonon scattering process, as indicated by the dash line in Fig. 1(a). This implies that the GB-scattering of charge carrier is weakened but not entirely removed in NbCoSn-Pt.

In contrast, NbCoSn-Pt-AN has a much higher $\mu_w$ than NbCoSn-Pt across the temperature range from 295 to 875 K, following a temperature dependence much closer to acoustic phonon-dominated scattering for single crystals. This enhancement in $\mu_w$ from NbCoSn-Pt to NbCoSn-Pt-AN cannot be ascribed to the different content of Pt as $m^*$ is independent of the Pt content, but should be due to the further weakened GB-scattering by annealing. This suggests that the impact of the microstructural evolution induced by post-SPS annealing is important for Pt-doped NbCoSn samples.

The weakened GB-scattering of charge carriers could also facilitate the phonon transport, which counterbalances the overall property improvement. A better evaluation of the effect of annealing on zT can be characterized by the thermoelectric quality factor $B$, which is proportional to the ratio of $\mu_w$ to the lattice thermal conductivity $\kappa_L$ [35]. The $B$ factor is an indicator of the best $zT$ that a material could achieve through optimal doping [4, 35]. The calculated $B$ was plotted as a function of temperature in Fig.1(b). Even though NbCoSn-Pt-AN has a slightly higher $\kappa_L$ than NbCoSn-Pt (Fig. S2 in Supplementary materials), the prominent enhancement in $\mu_w$ overbalanced the increase in $\kappa_L$, leading to an increased $B$ value for NbCoSn-Pt-AN. The increased $\mu_w$ and $B$ factor due to the weakened GB-scattering call for a thorough understanding of the local structural features of GBs and in turn to better understand the GB charge carrier transport.

*3.2 Microstructural investigation*

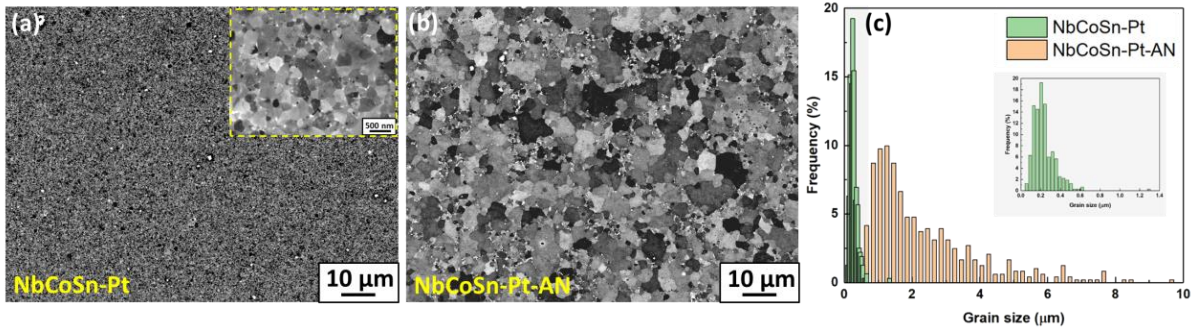

**Fig. 2.** Backscattered electron (BSE) images of (a) NbCoSn-Pt and (b) NbCoSn-Pt-AN. The inset in (a) is a magnified view showing the submicron grains of NbCoSn-Pt. (c) Grain size distribution of NbCoSn-Pt (green) and NbCoSn-Pt-AN (Orange).

During synthesis, the same BM and SPS parameters were used on both samples. BM is effective in producing nano-grained powders and SPS enables the fabrication of bulk nanocrystalline compounds [37]. The combination is often used to synthesize HH compounds [16, 37, 38]. The backscattered electron (BSE) images of NbCoSn-Pt and NbCoSn-Pt-AN are displayed in Fig. 2. The inset in Fig. 2(a) is a magnified view of NbCoSn-Pt, showing the submicron grains. The most striking difference between these two samples is the grain size: 0.23 ± 0.12 μm for NbCoSn-Pt, and 2.38 ± 1.65 μm for NbCoSn-Pt-AN. In addition, the grain size distribution is displayed in Fig. 2(c) for both samples. Post-SPS annealing hence leads to a tenfold increase in grain size and a widening of the size distribution in NbCoSn-Pt-AN. During the post-SPS annealing, the stored deformation energy induced by BM drives grain growth, leading to a largely reduced GB-area of NbCoSn-Pt-AN. The difference in grain size between these two samples becomes the first key factor that influences the $\mu_w$ through GB-scattering.

Near-atomically resolved APT analyses were performed on these Pt-doped samples to obtain local compositional information [39, 40]. Fig. 3(a) shows a representative atom map of the Pt distribution from the analysis of a specimen of NbCoSn-Pt. Four of five grains from the volume, labeled as G1, G2, G3 and G4, are visible along this observation direction. The GBs are highlighted by a strong enrichment of Pt atoms along GBs. The local excess of Pt at GBs is herein referred to as enrichment or segregation interchangeably. Interfacial energy, excess volume, grain boundary stress and diversity of atomic configurations at GBs often lead to chemical segregation, driven by the minimization of the system's free energy [41].

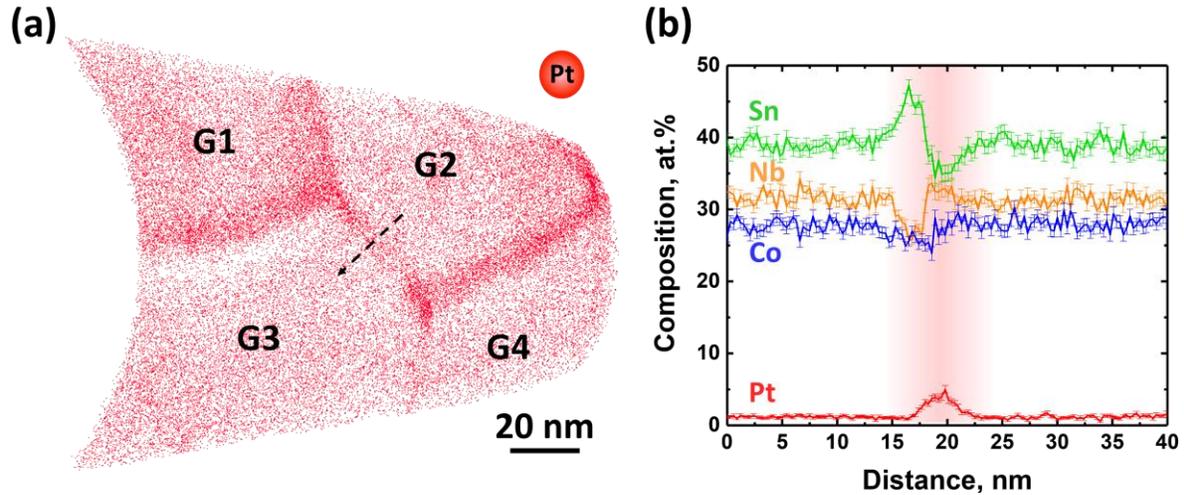

**Fig. 3.** APT analysis of NbCoSn-Pt: (a) distribution map of Pt atoms, indicating the presence of four grains (G1, G2, G3 and G4); (b) 1D-composition profile in a 20 nm-diameter cylinder perpendicular to the G2/G3 grain boundary.

1D-composition profiles along 20 nm-diameter cylinders positioned perpendicular to the GBs were calculated. Fig. 3(b) shows the 1D-composition profile across the G2/G3 grain boundary along the direction indicated by the arrow in Fig. 3(a). Composition profiles across other GBs are displayed in Fig. S3 (Supplementary materials). The peak Pt composition in NbCoSn-Pt is in the range of 3.4-12.8 at.%. Different GBs have a varying degree of Pt enrichment, which can be dependent on the geometric characteristics of the GB affected [42, 43]. The intragranular composition from different grains is summarized in Fig. 4, which shows a similar intragranular Pt-composition at $1.35 \pm 0.09$ at.%.

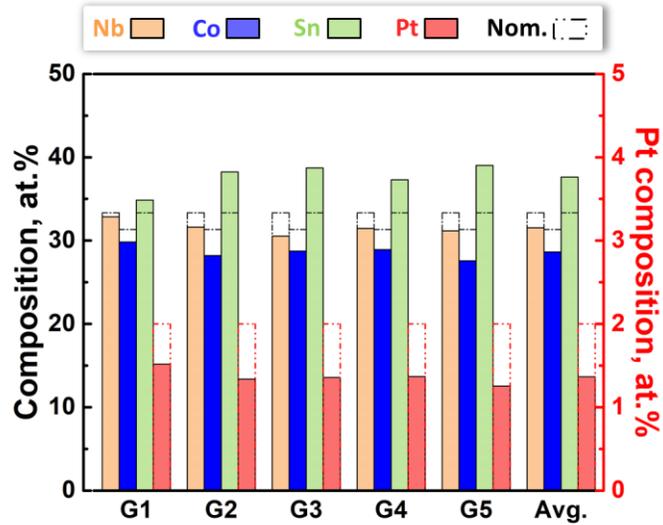

**Fig. 4.** Individual inner-grain composition of G1-G5 and the average inner-grain composition (Avg.), evaluated by APT for the volume shown in Fig. 3. The content of Nb, Co, Sn and Pt are present with orange, blue, green and red bars, respectively. The stoichiometric composition of the NbCoSn-Pt is indicated by the dash-dot bar.

Segregation of Pt to GBs is also observed in NbCoSn-Pt-AN. The sampled volume by APT is much smaller than the grain size of NbCoSn-Pt-AN (several micrometers (Fig. 2(b)), and APT datasets typically only contain one GB at most. Fig. 5(a) and (b) show the 1D-composition profiles across the GB of NbCoSn-Pt-AN in two APT datasets. The corresponding distribution map of Pt near GB is displayed in the inset. The maximum segregation content of Pt is ~10.7 at.% in Fig. 5(a) and ~6.3 at.% in Fig. 5(b). Importantly, the intragranular Pt-composition is approx. 1.27±0.09 at. %; markedly similar to the non-annealed NbCoSn-Pt sample.

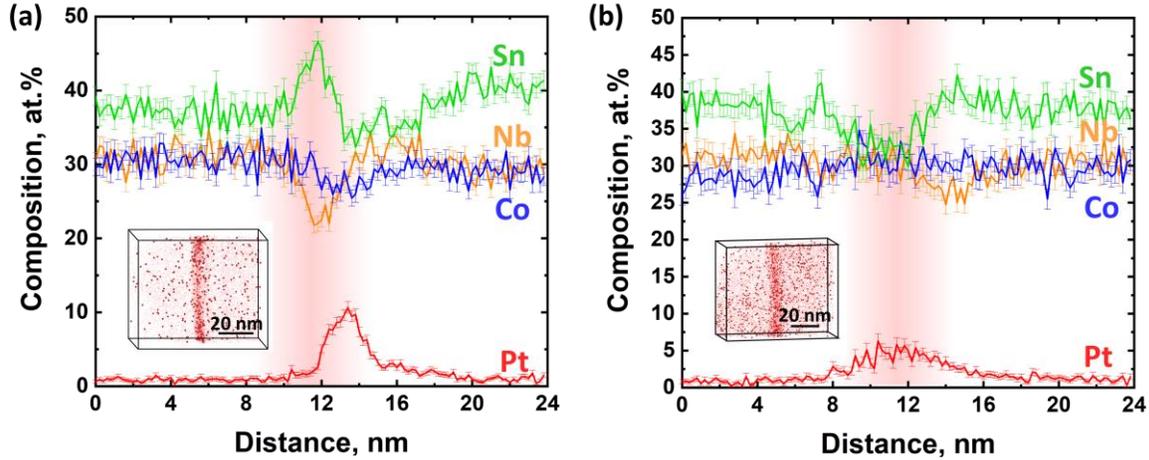

**Fig. 5.** (a) and (b) 1D-composition profiles across the GB of NbCoSn-Pt-AN in two APT datasets, with the inset showing the corresponding distribution map of Pt atoms.

Our APT analyses in Fig. 3 and Fig. 5 show that Pt dopants distribute within the matrix and segregate to GBs in both samples. In the grain interior, Pt dopants enhance the phonon scattering by point defects and thus decrease the lattice thermal conductivity. The average intragranular composition across multiple APT datasets is reported in Fig. 6. It shows that the intragranular Pt-content is within one standard deviation, i.e. ~1.35 ± 0.09 at.% for NbCoSn-Pt and 1.27 ± 0.09 at.% for NbCoSn-Pt-AN, which explains their similar Seebeck coefficients (Fig. S1(b), Supplementary materials), and suggests a Pt-solubility of approx. 1.3–1.4 at.% in NbCoSn. This further confirms that the increased $u_w$ from NbCoSn-Pt to NbCoSn-Pt-AN (Fig. 1(a)) at low temperatures could not be attributed to ionized impurity scattering as the two samples have similar intragranular content of Pt and thus similar concentration of ionized impurities. Rather, it supports the presence of GB-scattering of charge carriers. Similar phenomena have also been reported in $Mg_3Sb_2$ [9, 26] and other materials [36, 44, 45]. GB-scattering is not only dominant at low temperatures, but also affects high-temperature performance. In $Mg_3Sb_2$ [9, 26], even at high temperatures where the conductivity is finally decreasing with temperature, there is still sufficient GB resistivity to decrease $zT$ by ~20% or so.

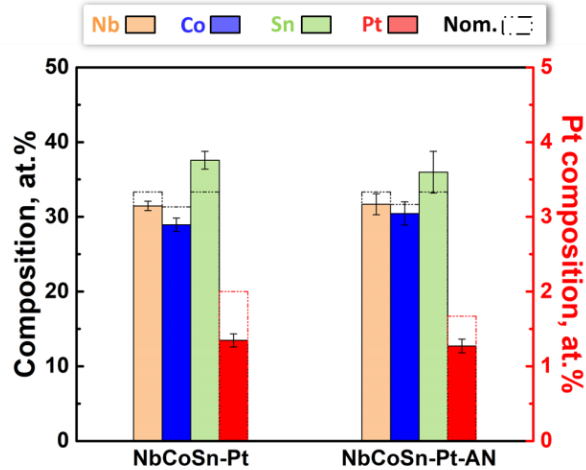

**Fig. 6.** A comparison in the intragranular composition of NbCoSn-Pt and NbCoSn-Pt-AN. The average content of Pt is similar: 1.35 ± 0.09 at.% for NbCoSn-Pt and 1.27 ± 0.09 at.% for NbCoSn-Pt-AN.

*3.3 In situ four-point-probe local electrical conductivity measurement*

To assess the impact of the segregation of Pt to GBs observed in both samples on the electrical conductivity, an *in situ* four-point-probe technique was used to measure the local electrical conductivity within a SEM [34]. This technique has sufficiently high sensitivity and spatial resolution to resolve the change in electrical conductivity across a single GB, even in a highly conductive material like Cu [34]. To enable these measurements, the grain size must be sufficiently large to separate the four probes, but neither of the two samples discussed above was suitable for this approach. We therefore selected the NbCoSn-Pt (*x*=0.05) sample before BM, with an average grain size of approx. 25 μm. The corresponding BSE image and EDS map of Pt are respectively shown in Fig. 7 (a) and (b), highlighting Pt enrichment and percolation along GBs.

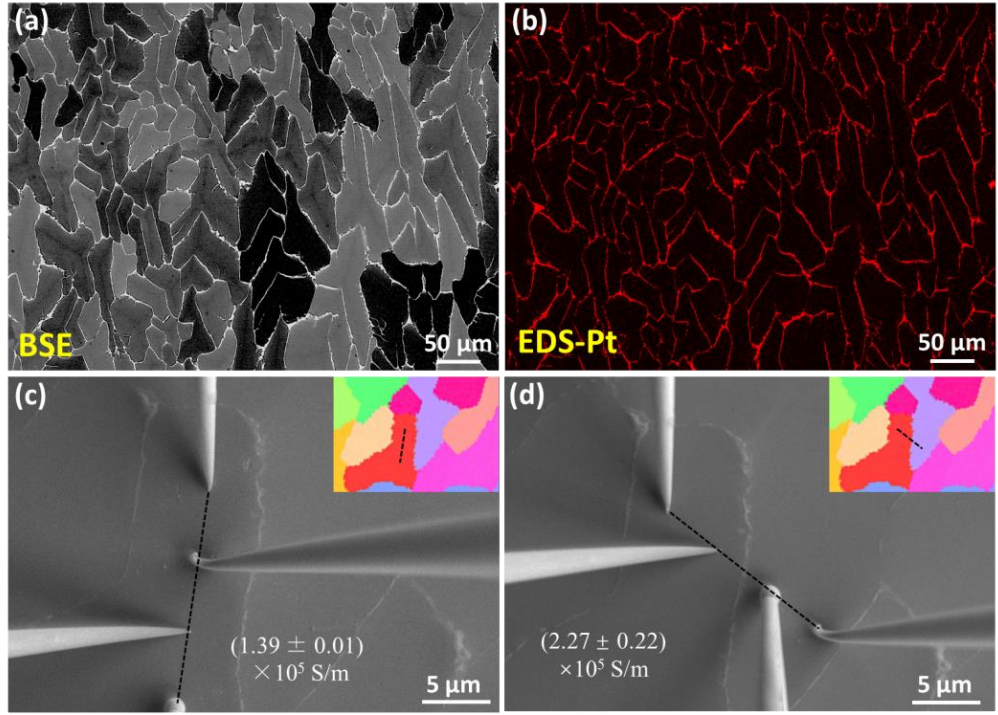

**Fig. 7.** (a) Backscattered electron (BSE) image of the pre-BM NbCo(Pt)Sn sample. b) The corresponding energy-dispersive X-ray spectroscopy (EDX) map of Pt. The white GBs in the BSE image is due to the enrichment of Pt. The position layout of the four needles used for the local measurement c) within the grain interior, and d) crossing a high angle grain boundary, which was assessed by electron-backscattered diffraction (inset).

The electrical conductivity was measured inside two adjacent grains and across the corresponding high angle grain boundary (HAGB), as assessed by electron-backscattered diffraction (inset in Fig. 7(c) and (d)). The local electrical conductivity was the average of five measurements, which were repeatedly conducted by relocating the needle positions within the region of interest. We also conducted the local measurement within the grain interior with different distances to GBs, and it shows that the electrical conductivity does not dependent on the distance to GBs. One example of the position layout of the four probes is displayed in Fig. 7(c) for grain interior and Fig. 7(d) for HAGB. The absolute value of the electrical conductivity is $(1.39 \pm 0.01) \times 10^5$ S/m for the grain interior, i.e. comparable with the value of NbCoSn-Pt-AN at room temperature, indicating that the electrical conductivity of the pre-BM sample and post-BM sample within a single grain are comparable and the local electrical conductivity values are reliable. Across the GB, and despite possible space-charge effects, the conductivity increases almost twofold to $(2.27 \pm 0.22) \times 10^5$ S/m thanks to the enrichment of Pt at GBs.

## 4. General discussion

*4.1 Microstructure-thermoelectric performance relationship*

We now wish to discuss the details of how the microstructure and the thermoelectric properties are linked, and their evolution through the annealing process.

First, a change in the properties associated to the post-SPS annealing could be associated to suppressed anti-site defects. It was reported that the anti-site defects could shrink the band gap, enhance the density of states slope near the Fermi level, and are favorable for the electrical transport properties of intrinsic semiconductors [46]. Therefore, if the impact of anti-site defects is significant, post-SPS annealing should decrease the electrical conductivity due to the decreased density of anti-site defects [46, 47]. However, an inverse trend is observed in our study: post-SPS annealing results in a higher electrical conductivity of NbCoSn-Pt-AN. In addition, X-ray diffraction (XRD) is an effective experimental method to examine the anti-site defects as they usually lead to different reflection positions or reflection intensities [48, 49]. However, as shown in our previous work [28], we did not observe a marked difference in the XRD patterns between $x$=0.6 (not annealed) and $x$=0.5 (annealed) sample. Therefore, the density of anti-site defects does not change significantly upon annealing and its impact is marginal in our study.

In contrast, the density of GBs is substantially different. GB-scattering affects the weighted mobility [35] of NbCoSn HH at low temperatures (Fig. 1(a)). The large difference in the weighted mobility of the pristine NbCoSn and NbCoSn-Pt demonstrates that the segregation of Pt to GBs lowers the GB scattering. Moreover, the scattering is further reduced by the microstructure evolution induced by post-SPS annealing, which modifies both the grain size and GB chemistry. We employed the two-phase model [26, 50] to interpret the impact of GB-associated features on the electrical conductivity. Due to the space-charge effect, GBs can trap charge carriers (e.g. electrons in n-type semiconductors), forming electron-depleted regions near GBs and leading to a high GB resistivity. In the two-phase model, GBs are considered as a GB-phase where an effective band offset ($\Delta E$) is applied compared to the neutral grains [26, 50]. The electrical conductivity of the GB-phase ($\sigma_{GB}$) is associated to $\Delta E$ through the relationship:

$$\sigma_{GB} = A \, exp(\frac{\Delta E}{k_B T}) \quad (1)$$

where A is a pre-expotinal factor, which is related to the charge carrier density $n$, the carrier effective mass $m^*$, Boltzmann constant $k_B$ and tempereature $T$ [50].

In a series circuit, the conductivity of the materials could be expressed as:

$$\sigma^{-1} = \sigma_G^{-1} + (\sigma_{GB}^{-1} - \sigma_G^{-1})t_{GB} \quad (2)$$

where $G$ and $GB$ refer to grain phase and grain boundary phase, respectively, while $t_{GB}$ is the size fraction of the GB-phase. This two-phase model accounts for the effect of grain size on the electrical conductivity through $t_{GB}$. However, this is insufficient to account for the local composition of GBs, e.g. the segregation of Pt as we observed herein. Compositional variation at GBs will have an impact on $\Delta E$ and thus influences $\sigma_{GB}$.

We provided direct experimental evidence that GB segregation, here of Pt, enhances the electrical conductivity, and it turns out that the space-charge effect of GBs could be partially or even entirely compensated by GB segregation. Similar experimental results have also been reported for ceria [51]. Therefore, both grain size and GB segregation influence the total electrical conductivity. The two-phase model needs to evolve to include the impact of GB segregation and a possible approach is to apply different $\Delta E$ for GBs (Equation 1) with different local compositions.

**Table 1.** Total electrical conductivity (σ), inner grain electrical conductivity ($\sigma_G$), the volume fraction of grain boundaries ($t_{GB}$) and the calculated electrical conductivity of grain boundaries ($\sigma_{GB}$) for NbCoSn-Pt and NbCoSn-Pt-AN.

|  | $\sigma$ (×10$^4$ S/m) | $\sigma_G$ (×10$^4$ S/m) | $t_{GB}$ (%) | $\sigma_{GB}$ (×10$^4$ S/m) |
|---|---|---|---|---|
| NbCoSn-Pt | 7.15 | 13.88 | 3.86 | 0.55 |
| NbCoSn-Pt-AN | 13.59 | 13.88 | 0.38 | 2.09 |

Based on both, local and bulk electrical transport measurements, as well as the grain size, we calculated $\sigma_{GB}$ in Equation 2 for NbCoSn-Pt and NbCoSn-Pt-AN. The solubility of Pt in NbCoSn imposes a similar intragranular composition of Pt for the pre-BM and post-BM samples. Therefore, $\sigma_G$ in Equation 2 was taken as the value of the electrical conductivity which was locally measured in the grain interior. The difference in the total electrical conductivity between NbCoSn-Pt and NbCoSn-Pt-AN is then attributed to the second term in Equation 2. We calculated the volume fraction of GBs by $t_{GB} = 1 - \left(\frac{d-h}{d}\right)^3$, where $d$ is the grain size and $h$ is the width of the GB-phase [52]. The data used in Equation 2 and the calculated $\sigma_{GB}$ are listed in Table 1. Here, the width of the GB-phase is assigned to 3 nm and its impact on the relative value of $\sigma_{GB}$ is minor. The calculated $\sigma_{GB}$ of NbCoSn-Pt-AN is approximately four times larger than that of NbCoSn-Pt.

In our study, the grain size and the content of Pt segregation at GBs are interrelated, due to the grain growth process between NbCoSn-Pt and NbCoSn-Pt-AN. The ten times larger grain size of NbCoSn-Pt-AN leads to a ~90% reduction in $t_{GB}$. In addition, grain growth also results in the intragranular Pt being repelled into GBs, since the content of Pt for both samples is higher than the estimated solubility of Pt in NbCoSn. The larger grain size of NbCoSn-Pt-AN is expected to push more Pt at GBs on average, leading to a higher value of $\sigma_{GB}$.

*4.2 Microstructural optimization*

Gaining a holistic understanding of the influence of the microstructure on TE performance is a necessary step to guide the microstructural optimization via, for instance, GB engineering, and improve the efficiency of future materials and devices. Our work demonstrates that grain size and GB segregation combine to influence the electrical conductivity. Due to the space-charge effects, pristine GBs are more resistive than the more neutral grains [26, 51], while the GB-electrical conductivity could be modified by GB segregation. Therefore, the effect of both factors must be considered to develop effective TE design strategies.

If the GB-electrical conductivity is lower than the grain interior, increasing the grain size would be helpful to enhance the total electrical conductivity. But the larger grain size could also lead to higher thermal conductivity [18, 22], so the optimization window through increasing the grain size might be narrow. Our work offers another potential strategy. As shown in the pre-BM sample, GB segregation is possible to partially or entirely compensate the space-charge effects associated with GBs. If this strategy could be employed in other TE materials, the electrical conductivity would be less sensitive to the grain size, and grain refinement might be favorable for the TE performance by reducing the thermal conductivity.

Ultimately, both the segregant and the magnitude of the segregation influence the GB-electrical conductivity [51, 53]. Therefore, thermodynamic (e.g. solubility of dopants, precipitation of secondary phases) and kinetics (e.g. grain growth and diffusion of dopants) should be considered in the design of future TE materials. Calculations of the solubility of dopants in the matrix and the corresponding electronic structure of GBs would further help optimize the microstructure. For HH compounds which are often synthesized through ball milling and sintering, grain growth and dopant diffusion are highly likely to happen in their service conditions at medium-to-high temperatures [10, 13, 14]. Therefore, approaches should be considered to stabilize the microstructure to improve the long-term performance of TE devices in service, e.g. optimizing the solute concentration [52, 54] or post-sintering annealing.

## 5. Conclusion

GB-scattering of charge carriers is the dominant scattering mechanism of the NbCo$_{1-x}$Pt$_x$Sn HH alloys, which deteriorates the weighted mobility of NbCoSn-Pt at low temperatures. After post-SPS annealing, NbCoSn-Pt-AN has a two times higher weighted mobility at room temperature compared to that of NbCoSn-Pt. We demonstrated that GB-scattering of charge carriers is influenced by the microstructure variation induced by post-SPS annealing because of a tenfold increase in grain size and Pt-segregation to GBs. The associated drop in electrical resistivity was confirmed by in-situ measurements. We hence demonstrate that dopant segregation, in our case Pt, can completely counteract the space-charge effect associated with GBs and thus leading to an overall higher GB-electrical conductivity. By adjusting the material composition and processing, it is hence possible to use grain boundary engineering to manipulate the GB-transport properties, opening new possibilities for optimizing TE performance.

## Acknowledgements

Dr. Ting Luo acknowledges the financial support from the Alexander von Humboldt Foundation. The authors are grateful to Christian Broß, Uwe Tezins and Andreas Sturm for their technical support of the atom probe tomography and focused ion beam facilities at Max-Planck-Institut für Eisenforschung GmbH. Dr. Federico Serrano-Sánchez acknowledges the funding from the European Union's Horizon 2020 research and innovation program under the Marie Skłodowska-Curie grant agreement (No. 839821). This work in Dresden was funded by the Deutsche Forschungsgemeinschaft (DFG, German Research Foundation)-Projektnummer (392228380). Dr. Hanna Bishara and Prof. Gerhard Dehm acknowledge the financial support by the ERC Advanced Grant GB CORRELATE (Grant Agreement 787446 GB-CORRELATE).

## Supplementary materials

Supplementary material associated with this article can be found, in the online version, at doi:

# Supplementary materials

**Dopant-segregation to grain boundaries assisted enhancement of electrical conductivity in n-Type NbCo(Pt)Sn half-Heusler alloy**

Ting Luo [a,*], Federico Serrano-Sánchez [b], Hanna Bishara [a], Siyuan Zhang [a], Ruben Bueno Villoro [a], Jimmy Jiahong Kuo [c], Claudia Felser [b], Christina Scheu [a], G. Jeffrey Snyder [c], James P. Best [a], Gerhard Dehm [a], Yuan Yu [d], Dierk Raabe [a], Chenguang Fu [b], Baptiste Gault [a, e, *]

[a] Max-Planck-Institut für Eisenforschung GmbH, Max-Planck Straße 1, 40237, Düsseldorf, Germany
[b] Max Planck Institute for Chemical Physics of Solids, Nöthnitzer Str. 40, 01187 Dresden, Germany
[c] Materials Science & Engineering (MSE), Northwestern University, Evanston, IL 60208, USA
[d] Institute of Physics IA, RWTH Aachen University, 52056 Aachen, Germany
[e] Department of Materials, Royal School of Mines, Imperial College, Prince Consort Road, London, SW7 2BP, UK

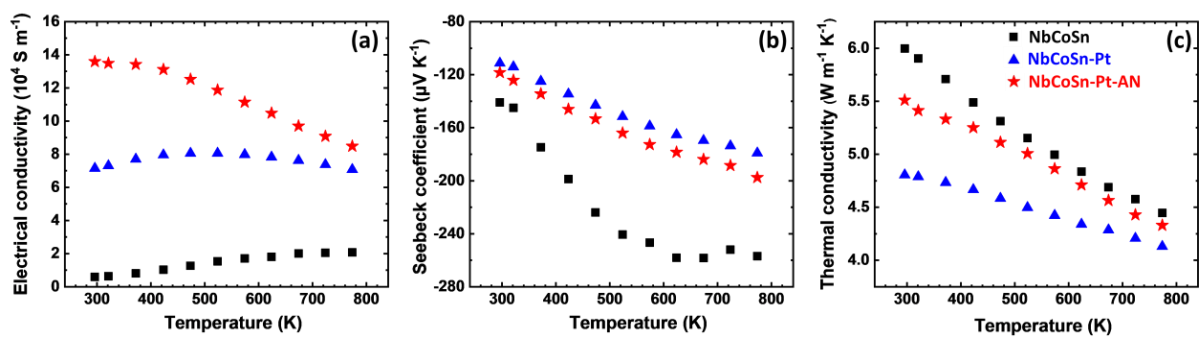

**Fig. S1.** Temperature dependence of (a) electrical conductivity, (b) Seebeck coefficient, and (c) thermal conductivity of NbCoSn, NbCoSn-Pt and NbCoSn-Pt-AN.[1]

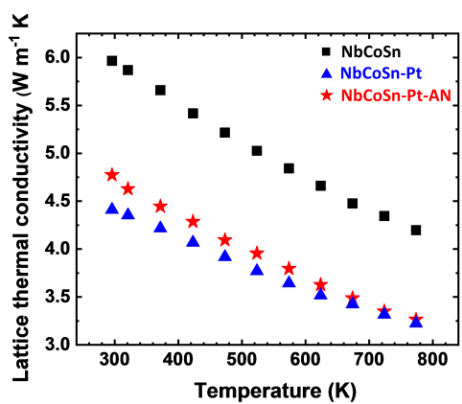

**Fig. S2.** Temperature dependence of lattice thermal conductivity of NbCoSn, NbCoSn-Pt and NbCoSn-Pt-AN.[1]

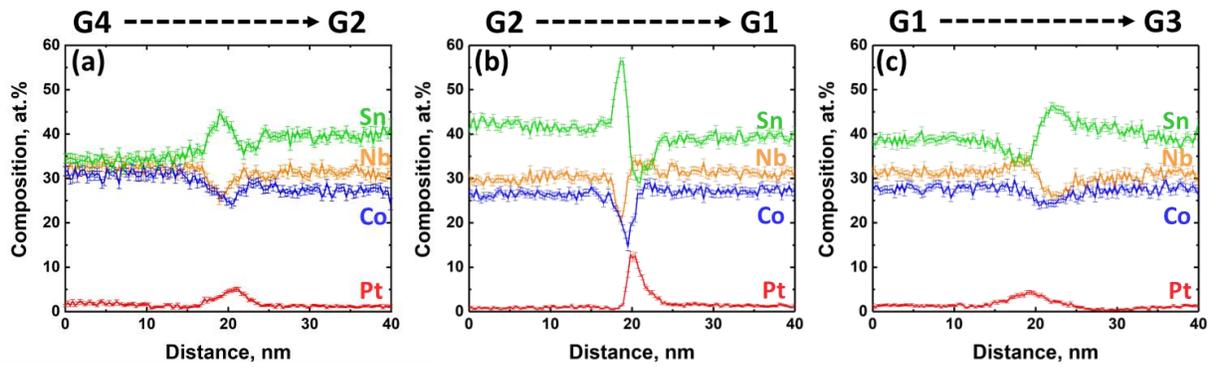

**Fig. S3.** 1D-composition profiles in 20nm-diameter cylinders perpendicular to (a) the G4/G2 grain boundary, (b) the G2/G1 grain boundary and (c) the G1/G3 grain boundary. G1, G2, G3 and G4 grains are displayed in the APT volume in Figure 3a.